\documentclass[a4paper,superscriptaddress,prb,showpacs]{revtex4-1}
\usepackage{graphicx}
\usepackage{amssymb}

\begin{document}

\title{Phase sensitive Landau-Zener-St\"uckelberg interference in superconducting quantum circuit}

\author{Zhi-Xuan Yang}
\affiliation{Shenzhen Institute for Quantum Science and Engineering, \\
Southern University of Science and Technology, Shenzhen, Guangdong, China}

\author{Yi-Meng Zhang}
\affiliation{Shenzhen Institute for Quantum Science and Engineering, \\
Southern University of Science and Technology, Shenzhen, Guangdong, China}

\author{Yu-Xuan Zhou}
\affiliation{Shenzhen Institute for Quantum Science and Engineering, \\
Southern University of Science and Technology, Shenzhen, Guangdong, China}
\affiliation{Department of Physics, Southern University of Science and Technology,\\
Shenzhen, Guangdong, China}

\author{Li-Bo Zhang}
\affiliation{Shenzhen Institute for Quantum Science and Engineering, \\
Southern University of Science and Technology, Shenzhen, Guangdong, China}
\affiliation{Guangdong Provincial Key Laboratory of Quantum Science and Engineering, \\
Southern University of Science and Technology, Shenzhen, Guangdong, China}
\affiliation{Shenzhen Key Laboratory of Quantum Science and Engineering, \\
Southern University of Science and Technology, Shenzhen, Guangdong, China}

\author{Fei Yan}
\affiliation{Shenzhen Institute for Quantum Science and Engineering, \\
Southern University of Science and Technology, Shenzhen, Guangdong, China}
\affiliation{Guangdong Provincial Key Laboratory of Quantum Science and Engineering, \\
Southern University of Science and Technology, Shenzhen, Guangdong, China}
\affiliation{Shenzhen Key Laboratory of Quantum Science and Engineering, \\
Southern University of Science and Technology, Shenzhen, Guangdong, China}

\author{Song Liu}
\affiliation{Shenzhen Institute for Quantum Science and Engineering, \\
Southern University of Science and Technology, Shenzhen, Guangdong, China}
\affiliation{Guangdong Provincial Key Laboratory of Quantum Science and Engineering, \\
Southern University of Science and Technology, Shenzhen, Guangdong, China}
\affiliation{Shenzhen Key Laboratory of Quantum Science and Engineering, \\
Southern University of Science and Technology, Shenzhen, Guangdong, China}

\author{Yuan Xu}
\affiliation{Shenzhen Institute for Quantum Science and Engineering, \\
Southern University of Science and Technology, Shenzhen, Guangdong, China}
\affiliation{Guangdong Provincial Key Laboratory of Quantum Science and Engineering, \\
Southern University of Science and Technology, Shenzhen, Guangdong, China}
\affiliation{Shenzhen Key Laboratory of Quantum Science and Engineering, \\
Southern University of Science and Technology, Shenzhen, Guangdong, China}

\author{Jian Li} \email{lij33@sustech.edu.cn}
\affiliation{Shenzhen Institute for Quantum Science and Engineering, \\
Southern University of Science and Technology, Shenzhen, Guangdong, China}
\affiliation{Guangdong Provincial Key Laboratory of Quantum Science and Engineering, \\
Southern University of Science and Technology, Shenzhen, Guangdong, China}
\affiliation{Shenzhen Key Laboratory of Quantum Science and Engineering, \\
Southern University of Science and Technology, Shenzhen, Guangdong, China}

\begin{abstract}
Superconducting circuit quantum electrodynamics (QED) architecture composed of superconducting qubit and resonator is a powerful platform for exploring quantum physics and quantum information processing. By employing techniques developed for superconducting quantum computing, we experimentally investigate phase-sensitive Landau-Zener-St\"uckelberg (LZS) interference phenomena in a circuit QED. Our experiments cover a large range of LZS transition parameters, and demonstrate the LZS induced Rabi-like oscillation as well as phase-dependent steady-state population.
\end{abstract}


\maketitle

\section{Introduction}

Landau-Zener-St\"uckelberg (LZS) interference is a phenomenon that appears in a two-level quantum system undergoing periodic transitions between energy levels at the anticross.\cite{nori_review} Since the pioneering theoretical work by Landau,\cite{landau1, landau2} Zener,\cite{zener1} St\"uckelberg,\cite{stuckelberg1} and Majorana \cite{majorana1} in 1932, in the past nearly 90 years profound explorations on LZS interference and related problems have been carried out both theoretically and experimentally. Especially entered the 21st century, LZS interference and related researches become active again, thanks to the enormous progresses in experimental technologies for creating and manipulating coherent quantum states in solid-state systems. LZS interference patterns have been observed in electronic spin of a double quantum dot,\cite{petta1} charge oscillations of a double quantum dot,\cite{ota1} electronic spin of nitrogen-vacancy (NV) center in a diamond,\cite{zhou14} and superconducting qubits.\cite{oliver1, sillanpaa1, wilson1, sun1} Recently, a method of using light propagating along the curved waveguides to simulate the time domain Landau-Zener transitions and LZS interferences has been proposed;\cite{liu1} even in a hybrid system formed by a superconducting qubit coupled to an acoustic resonator, LZS interference has been studied.\cite{sillanpaa2}

Unlike in double quantum dots and NV centers natural microscopic particles such as electron and atom being used to perform experiments, superconducting qubits \cite{devoret_review, oliver_review} consisting of Josephson junctions, inductors and capacitors are true artificial macroscopic quantum objects. The macroscopic nature of superconducting qubits makes them not only an ideal testbed for quantum physics but also an advanced platform for emerging quantum information processing.\cite{wendin_review} In early stage of LZS interference experiments in superconducting qubits,\cite{oliver1, sillanpaa1, wilson1, sun1} flux qubit\cite{mooij1}, charge qubit\cite{tsai1} and phase qubit\cite{martinis1} have been used. These qubits have fairly short coherence time of only a few nanoseconds, which makes observation of LZS dynamics, particularly the global structures\cite{garraway1} (also called the Rabi-like oscillations\cite{zhou14, LZSM_lasing}) which strongly dependent on the initial phase of periodic transitions hardly possible.

A new type of charge-noise-insensitive superconducting qubit called transmon has been invented in 2007.\cite{koch07} By combining transmon with superconducting coplanar waveguide (CPW) resonator, a superconducting quantum circuit working in microwave regime and equivalent to optical cavity quantum electrodynamics (QED) can be formed, which is known as circuit QED.\cite{wallraff04} In a circuit QED, the CPW resonator can not only protect the qubit from spontaneous decay, but also provide high-visibility readout of qubit states. Owing to refinements in circuit design, material platform and fabrication techniques in the past decade, circuit QED with transmon and its decendent Xmon\cite{barends1} becomes the mainstream of superconducting quantum information processing.\cite{google_supremacy} Nowadays the coherence time of a transmon or Xmon in a circuit QED system can easily be over $10\ \mu$s, which makes it an excellent system for studying LZS dynamics.

In this paper we adopt a circuit QED system consisting of an Xmon qubit, a transverse ($XY$) control line, a longitudinal ($Z$) control line and a readout resonator to investigate LZS interference phenomena. Combining good coherence of the Xmon qubit and controlling of the initial phase of periodic transitions, we demonstrate phase sensitivity of the qubit population in both time evolutions and steady states, and observe the Rabi-like oscillation for the first time in a superconducting quantum circuit. Since in our work the LZS interference does not happen in Schr\"odinger picture, one may consider our system as a quantum simulator to mimic LZS physics in a rotating frame.

\section{Physical model}

The equivalent diagram of the circuit QED system is shown in the dashed box of Fig.~\ref{fig:setup}. The dual-junction Xmon qubit can be excited by a near-resonance microwave driving field sending to the $XY$ control line. The qubit transition frequency can be tuned by magnetic flux threading the loop formed by the two Josephson junctions (the so-called $dc$ SQUID), and the magnetic flux is controlled by sending a current to the $Z$ control line. This tunability of qubit transition frequency gives rise to the periodic transitions required by LZS process.

For an Xmon qubit, the flux dependent transition frequency is approximately
\begin{equation}
\omega_{10}(\Phi)\approx \left. \left[ \sqrt{8E_CE_J|\cos(\pi\Phi/\Phi_0)|} - E_C \right] \right/ \hbar. \label{eq:omega_10}
\end{equation}
Here $\Phi$ is the flux threading Xmon's $dc$ SQUID loop, $\Phi_0 = h/(2e)$ is the magnetic flux quantum, $E_C$ and $E_J$ represent the single-electron charging energy and the total Josephson energy, respectively. From Eq.~(\ref{eq:omega_10}), one can easily find out that the qubit transition frequency is not linearly dependent on flux. Only when $\Phi/\Phi_0$ is far away from $\kappa/2$ (here $\kappa$ is an arbitrary integer), $\omega_{10}$ is near linearly dependent on $\Phi$. If the total flux $\Phi$ consists of a $dc$ part and an $ac$ part $\Phi = \Phi_{dc}+\Phi_{ac}\cos(\omega t + \phi)$, and $(\Phi_{dc} \pm \Phi_{ac})/\Phi_0$ is far away from $\kappa\pi/2$ (we may call the region far away from $\kappa\pi/2$ the near-linear region), the qubit transition frequency then becomes
\begin{equation}
\omega_{10}(\Phi) \approx \omega_0 - A\cos(\omega t + \phi),
\label{eq:omega_10_approx}
\end{equation}
with $\omega_0 \approx \omega_p - E_C / \hbar$ the $dc$ transition frequency, $A\approx \pi \omega_p \Phi_{ac} \tan(\pi\Phi_{dc}/\Phi_0)/(2\Phi_0)$ the transition frequency ($Z$) modulation amplitude, and $\omega$ the modulation frequency. Here $\omega_p \equiv\sqrt{8E_CE_J|\cos(\pi\Phi_{dc}/\Phi_0)|} / \hbar$ is called the plasma oscillation frequency. A detailed derivation of Eq.~(\ref{eq:omega_10_approx}) can be found in Appendix A.

The Hamiltonian of an Xmon qubit flux biased in near-linear region, with $XY$ microwave drive, can then be written as
\begin{equation}
\hat{H} = -\frac{\hbar}{2} \left[ \omega_0 - A\cos(\omega t + \phi) \right] \hat\sigma_z + \Omega\cos(\omega_d t)\hat\sigma_x ,
\label{eq:H_SP}
\end{equation}
with $XY$ drive frequency $\omega_d$, and Rabi frequency $\Omega$. For $XY$ drive, a fixed phase $0$ is assumed, so $\phi$ represents both the initial phase of periodic transitions ($Z$ modulation) and the relative phase between $Z$ modulation and $XY$ drive.

Bringing the Hamiltonian Eq.~(\ref{eq:H_SP}) to a rotating frame with frequency $\omega_d$, by considering only near resonance drive, $|\omega_d - \omega_0|\ll \omega_0$, we can make a rotating wave approximation (RWA) to ignore fast rotating terms with $e^{\pm 2i\omega_d t}$, and have
\begin{equation}
\hat{\widetilde{H}} = -\frac{\hbar}{2}\left[ \varepsilon_0 - A\cos(\omega t + \phi) \right]\hat\sigma_z + \frac{\hbar}{2}\Omega \hat\sigma_x ,
\end{equation}
where $\varepsilon_0 = \omega_0 - \omega_d$ is the detuning between qubit transition and $XY$ drive. This rotating-frame Hamiltonian is very similar to the LZS models being studied in literatures such as Ref.~\cite{nori_review} and Ref.~\cite{garraway1}. Taking relaxation and dephasing into account, time evolution of this system can be calculated by solving the Markov master equation \cite{carmichael91}
\begin{equation}
\dot{\hat\rho} = -\frac{i}{\hbar}\left[ \hat{\widetilde{H}} , \hat\rho \right] + \frac{\Gamma_1}{2}\left( 2\hat\sigma_- \hat\rho \hat\sigma_+ - \hat\sigma_+ \hat\sigma_- \hat\rho - \hat\rho \hat\sigma_+ \hat\sigma_- \right) + \frac{\Gamma_\varphi}{2}\left( \hat\sigma_z \hat\rho \hat\sigma_z - \hat\rho \right) ,
\label{eq:master_eq}
\end{equation}
where $\hat\rho$ is qubit density operator, $\hat\sigma_{+(-)}$ is qubit raising (lowering) operator, $\Gamma_1$ is energy relaxation rate, and $\Gamma_\varphi$ is pure dephasing rate.

\section{Experimental setup}

In previous experiments like in Ref.~\cite{my1, LZSM_lasing, my2}, continuous microwaves were used for $Z$ modulation and $XY$ drive (if needed). By using continuous microwaves, only steady-state properties of the systems can be studied. Since the relative phase $\phi$ is not locked during continuous-wave measurements, the steady states are actually averaged over $\phi$, therefore, any phase-sensitive phenomenon is washed away by this phase averaging.

\begin{figure}[h]
\includegraphics[width=14cm]{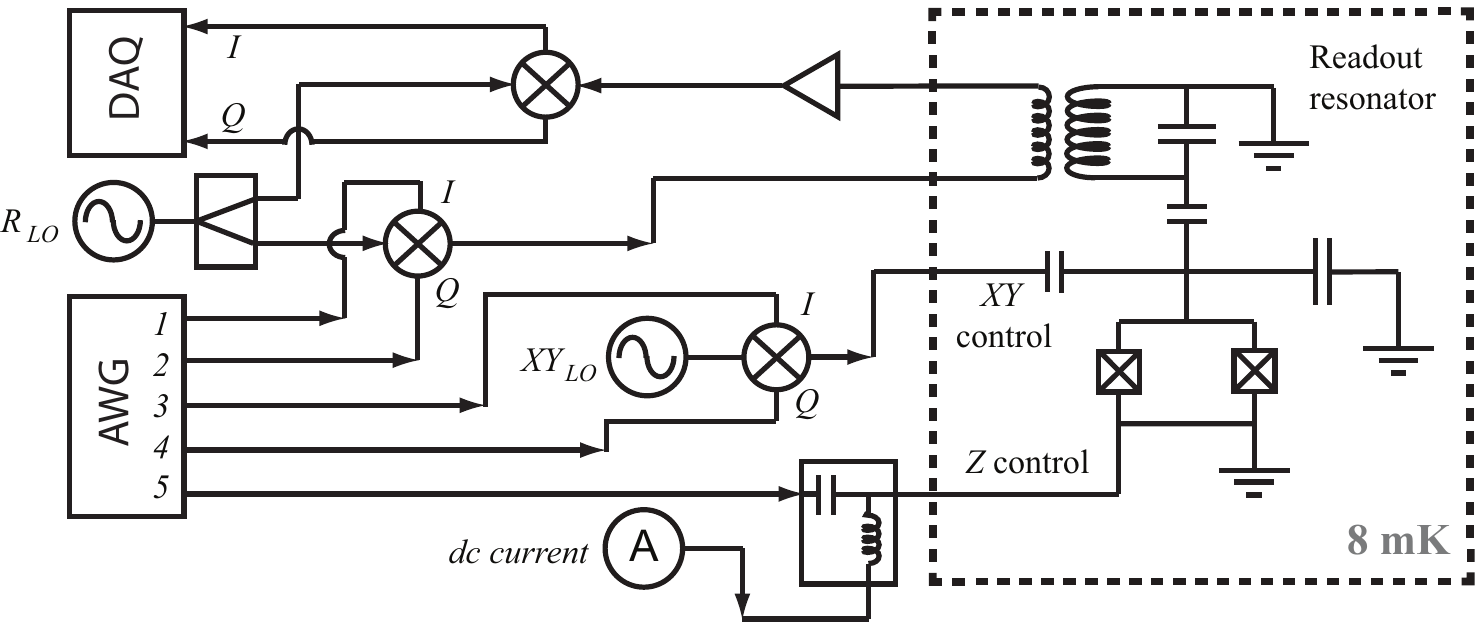}
\caption{
The schematics of experimental setup. The circuit QED sample consisting of an Xmon qubit with $XY$, $Z$ control lines and a readout resonator is mounted to the mixing chamber (illustrated by the dashed box) of a dilution refrigerator. The room temperature equipment and components are shown in the left-hand-side of this figure.}
\label{fig:setup}
\end{figure}

In order to obtain phase-sensitive evolutions, we need to precisely control the relative phase $\phi$. In our experiments, we use pulsed $Z$ modulation and standard $IQ$ mixing technique for $XY$ drive pulses to achieve phase control. As shown in Fig.~\ref{fig:setup}, the qubit $XY$ drive is formed by mixing a continuous local oscillator (LO) microwave field ($XY_{LO}$), with two pulsed quadratures ($I$ and $Q$) generated by two channels of an arbitrary waveform generator (AWG), $\omega_d = \omega_{LO} + \omega_{IQ}$. The pulsed $Z$ modulation signal is directly generated by another channel of the AWG. Since all channels of the AWG are synchronized, control $\phi$ is simply control the relative phase between signals from different channels. The $Z$ modulation signal and a $dc$ current for $dc$ flux bias are combined by a bias-tee and sent to the $Z$ control line of the Xmon.

\begin{figure}[h]
\includegraphics[width=7.5cm]{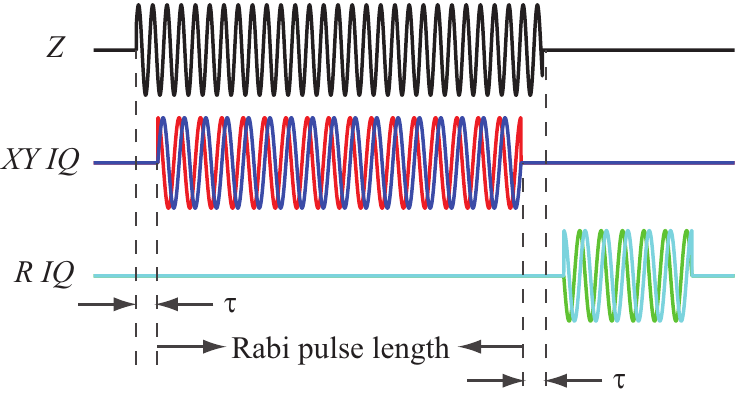}
\caption{
Schematic pulse sequences of one measurement cycle. $Z$ modulation pulse is a simple cosine function $\cos(\omega t + \phi)$, and it is $2\tau$ longer than $XY\ IQ$ pulses, $\cos(\omega_{IQ}t)$ and $\sin(\omega_{IQ}t)$ which define the Rabi pulse length and amplitude.}
\label{fig:pulse}
\end{figure}

The qubit dispersive readout \cite{wallraff05} pulses are generated in the same way as for $XY$ drive pulses. The readout ($R$) pulses through the readout resonator, are amplified at both low temperature and room temperature, then are down-mixed (a reverse process of $IQ$ mixing) and digitized by a data acquisition (DAQ) card. An illustration of pulse sequences is shown in Fig.~\ref{fig:pulse}. One should note that, due to different cable lengths as well as different components on cables, even though pulses are generated by the AWG at the same time, there are constant delays between $Z$, $XY$ and $R$ pulses when they reach the circuit QED sample at mixing chamber of the dilution refrigerator. To compensate these delays, we let the $Z$ modulation pulse a few tens of nanoseconds longer than the $XY\ IQ$ pulses.

\section{Results}

As pointed out by Garraway and Vitanov \cite{garraway1} that in small coupling regime $\Omega \ll \omega$, $\phi$ does not affect qubit population evolution substantially, thus in the following studies we focus on strong coupling regime $\Omega \gtrsim \omega$. However, the small coupling regime is useful for estimating the $Z$ modulation amplitude $A$ which can not be directly measured, since qubit excited state population follows the Bessel function of the first kind $J_n(A/\omega)$. Figure~\ref{fig:calib}(b) in Appendix B shows a mapping from $Z$ pulse peak-to-peak voltage to $A$ estimated in small coupling regime.

Depending on how large $A$ is, adiabatic limit $\Omega^2 \gg A\omega$ and non-adiabatic limit $\Omega^2 \ll A\omega$ can be reached. It is worth to point out that, in literatures like Ref.~\cite{nori_review}, slightly different definitions of parameter regimes are used. Slow-passage limit and fast-passage limit are defined as $\delta = \Omega^2/(4v) \gg 1$ and $\delta \ll 1$, respectively, where $v = A\omega \sqrt{1-(\varepsilon_0/A)^2}$. For on-resonance $XY$ drive ($\varepsilon_0 = 0$), slow-passage limit becomes $\Omega^2/(4A\omega) \gg 1$, which is equivalent to adiabatic limit. We choose to use adiabatic and non-adiabatic limits. By numerical simulations, we find that in non-adiabatic limit the qubit population evolution is only weakly dependent on $\phi$, therefore, we restrict our experiments in adiabatic limit as well as on the border of the two limits $\Omega^2 \approx A\omega$. Readers who are interested in LZS interference experiments in non-adiabatic limit in a transmon-type superconducting qubit, can check Refs. \cite{my1,my2}.

In the following LZS interference experiments, we $dc$ flux bias the qubit to frequency $\omega_0/2\pi = 4.365$ GHz, which is in the near-linear region, as indicated by the yellow arrow in Fig.~\ref{fig:qubit_spectro} of Appendix B. The measured energy relaxation time at this bias point is $T_1 = 1/\Gamma_1 = 26.4\ \mu$s, and the measured Ramsey dephasing time is $T_2 = 625$ ns, which is short due to biasing far away from the optimal point ($\Phi/\Phi_0 = 0$). However, one should note that since there is an $XY$ drive, the relevant pure dephasing rate should be the driven one, $\Gamma_\varphi$, which can be extracted from fitting the decay (with a rate approximately $\frac{3}{4}\Gamma_1 + \frac{1}{2}\Gamma_\varphi$) of Rabi oscillations. The driven pure dephasing rate varies slightly with Rabi frequency, thus $\Gamma_\varphi \approx 0.18$ MHz will be taken for later numerical calculations. This driven dephasing rate is small enough to allow coherent interference to exist in a long sequence of LZS process, and large enough to make sure that a reasonable length (40 $\mu s$) Rabi pulse can drive the qubit to a steady state.

\subsection{Adiabatic limit}

\begin{figure}[h]
\includegraphics[width=14cm]{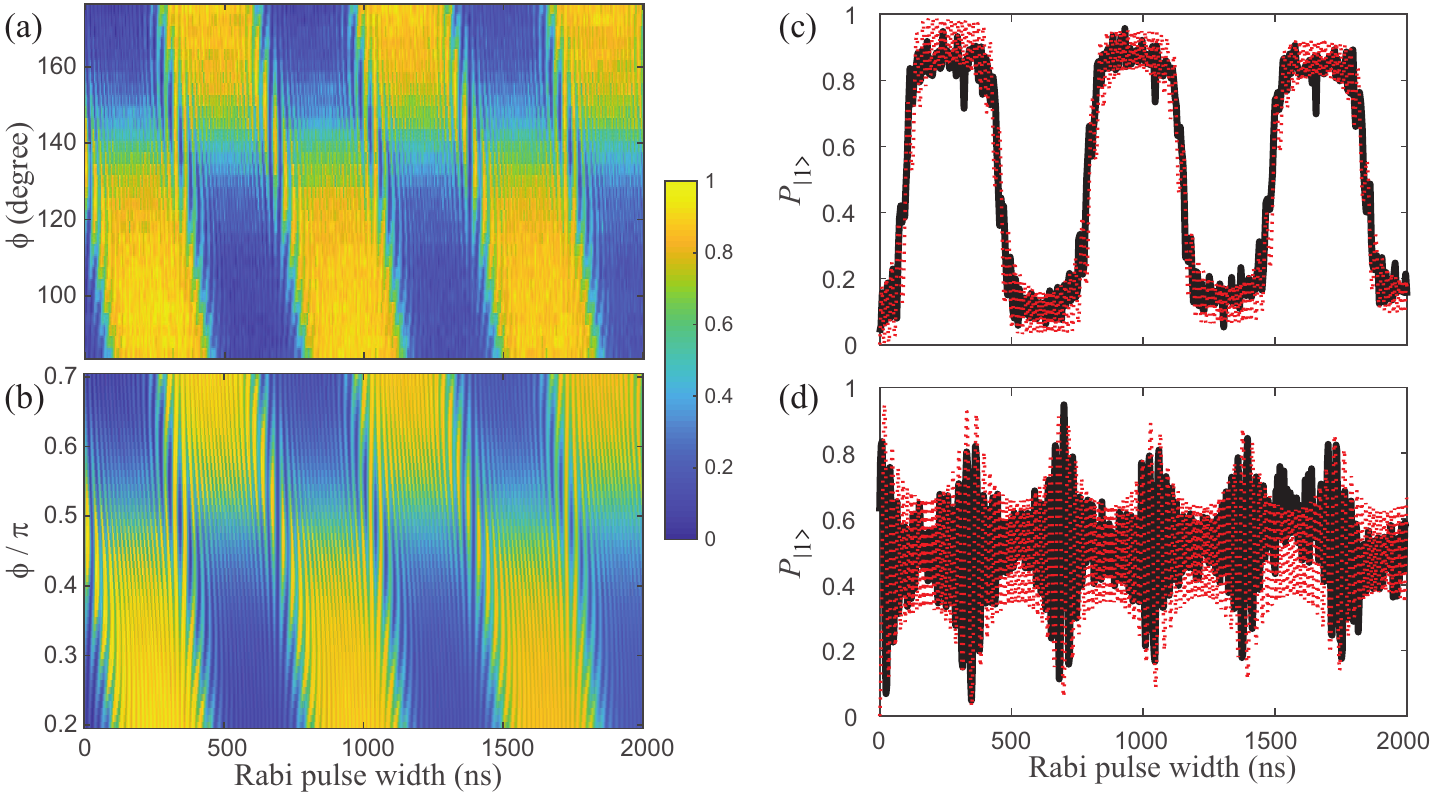}
\caption{
Evolution of $P_{|1\rangle}$, with $A/\omega = 50$ and $\Omega/\omega = 18.2$. (a) Experimental data for different relative phase $\phi$. (b) Numerical results corresponding to (a). (c) and (d) are $P_{|1\rangle}$ time evolutions for $\phi = 85$ degree ($0.2\pi$) and $\phi = 139$ degree ($0.5\pi$), respectively. The black solid curves are experimental data, and the red dotted curves are numerical results.}
\label{fig:adiabatic}
\end{figure}

We first study the phase-dependent time evolution of qubit excited state ($|1\rangle$) population $P_{|1\rangle}$ in adiabatic limit. $\Omega /2\pi = 26.2$ MHz, $\omega/2\pi = 1.44$ MHz, $A/2\pi = 72$ MHz, and $\varepsilon_0 = 0$ are taken. These values are close to those listed under Fig.~3 of Ref.~\cite{garraway1}. The relative phase $\phi$ is swept from 85 degree to 175 degree, and the measured results are shown in Fig.~\ref{fig:adiabatic}(a). Numerical results by solving the master equation Eq.~(\ref{eq:master_eq}) are shown in Fig.~\ref{fig:adiabatic}(b). One may find that $\phi$ in numerical simulation is from $0.2\pi$ to $0.7\pi$. This is due to the delay between $Z$ and $XY$ pulses, as discussed in Section 3. To avoid confusing the experimental and numerical values of $\phi$, we use numerical $\phi$ (in $\pi$) for the following discussions.

From both experimental and numerical results, we can find that the population evolution around $\phi = 0.5\pi$ differs from other $\phi$ values drastically. In Fig.~\ref{fig:adiabatic}(c) and (d), we plot population evolution for $\phi = 0.2\pi$ and $\phi = 0.5\pi$, respectively. Both experimental (black solid curves) and numerical (red dotted curves) results match Garraway and Vitanov's prediction \cite{garraway1} well.

\subsection{Boundary between adiabatic and non-adiabatic limits}

\begin{figure}[h]
\includegraphics[width=14cm]{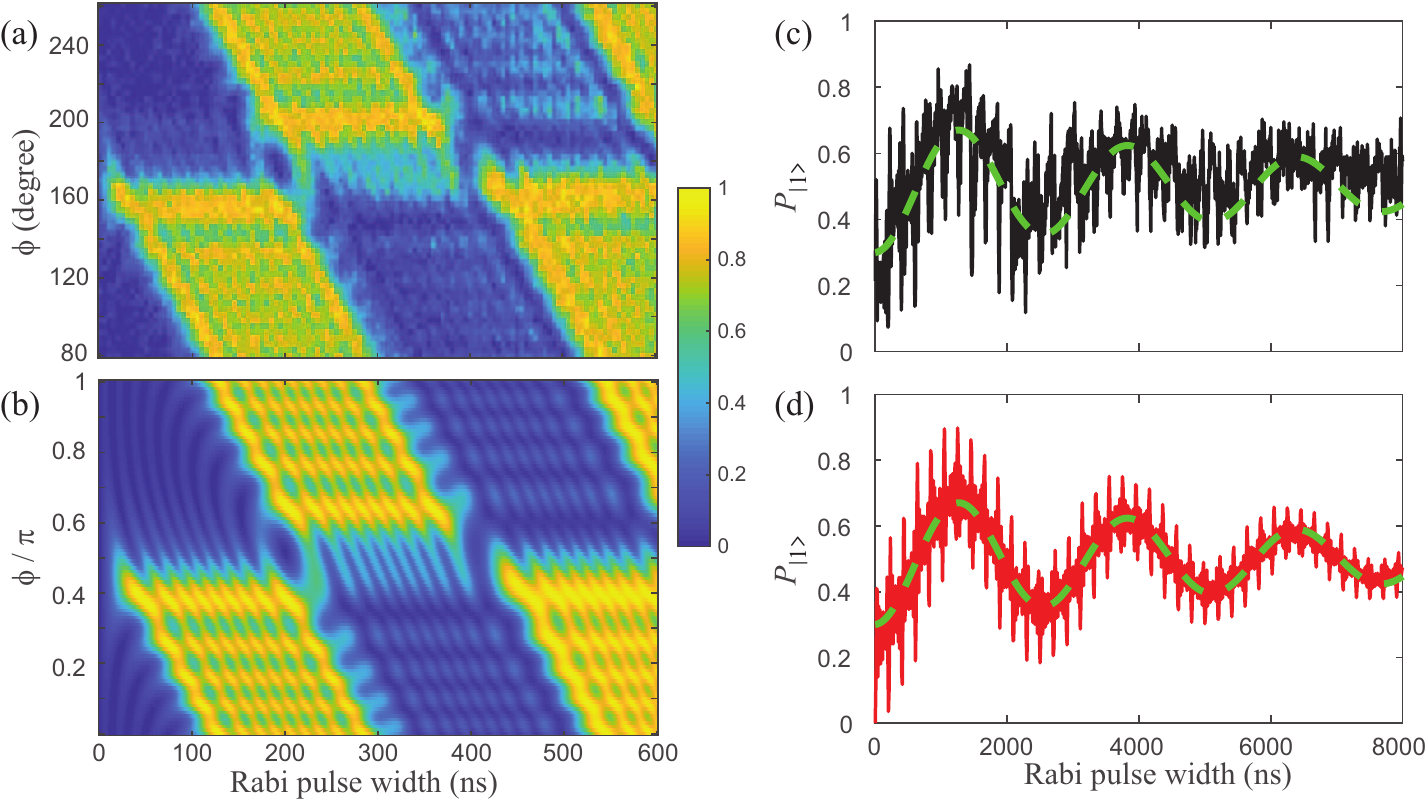}
\caption{
Evolution of $P_{|1\rangle}$, with $A/\omega = 26.6$ and $\Omega/\omega = 5$. (a) Experimental data for different relative phase $\phi$. (b) Numerical results corresponding to (a). (c) and (d) are experimental and numerical $P_{|1\rangle}$ time evolutions for $\phi = 176$ degree ($0.5\pi$), respectively. The green dashed curves in (c) and (d) are the same, indicating a Rabi-like oscillation of frequency $\Omega_0 = 2\pi\times390$ kHz.}
\label{fig:boundary_strong}
\end{figure}

We then study the phase-dependent time evolution of $P_{|1\rangle}$ at the boundary between adiabatic and non-adiabatic limits, $\Omega^2 \approx A\omega$. $\Omega /2\pi = 12$ MHz, $\omega/2\pi = 2.4$ MHz, $A/2\pi = 63.75$ MHz, and $\varepsilon_0 = 0$ are taken. The measured results and the corresponding numerical calculation results are shown in Fig.~\ref{fig:boundary_strong}(a) and (b), respectively. Similar to adiabatic limit, one can also find that the population evolution around $\phi = 0.5\pi$ differs from other $\phi$ values drastically. A longer trace of $P_{|1\rangle}$ evolution at $\phi = 0.5\pi$ is recorded, as indicated by the black solid curve in Fig.~\ref{fig:boundary_strong}(c). In this longer evolution trace a Rabi-like oscillation is observed, with a oscillation frequency $\Omega_0$ much lower than any of $\Omega,\ A$ and $\omega$. The numerical results for $\phi = 0.5\pi$ trace is shown in Fig.~\ref{fig:boundary_strong}(d) as the red solid curve.

In Ref.~\cite{LZSM_lasing}, a formula for Rabi-like oscillation frequency
\begin{equation}
\Omega_0 = 2\omega\sqrt{P_{LZ}}/\pi ,
\end{equation}
was derived, where $P_{LZ} = \exp(-2\pi\delta)$ is the Landau-Zener probability, and $\delta = \Omega^2/(4A\omega)$ as mentioned before. By putting the experimental values of $\Omega,\ A$ and $\omega$ into this formula, we get $\Omega_0\approx 2\pi \times 730$ kHz, which differs from the fitting (the green dashed curves in Fig.~\ref{fig:boundary_strong}) value $\Omega_0 = 2\pi \times 390$ kHz by roughly a factor of 2. Another formula
\begin{equation}
\Omega_0 = \omega\beta/\pi ,
\end{equation}
can be found in Eq.~(28) of Ref.~\cite{garraway1}, with
\begin{eqnarray}
&& \beta = \arccos(1-2P), \nonumber \\
&& P = \sin^2(\chi) \sin^2(\phi_{LZ} + 2\phi_{ad}), \nonumber \\
&& \chi = \arccos[\exp(-\pi\alpha^2/2)], \nonumber \\
&& \phi_{LZ} = \arg \Gamma(1-i\alpha^2/2) + \frac{\pi}{4} + \frac{\alpha^2}{2}\left[\ln\left( \frac{\alpha^2}{2} \right) -1 \right] , \nonumber \\
&& \alpha = \Omega / \sqrt{2A\omega}, \nonumber \\
&& \phi_{ad} = \sqrt{\Omega^2 + A^2} E(\cos\Theta)/(2\omega) , \nonumber \\
&& {\rm and} \ \ \Theta = \arctan(\Omega/A) , \nonumber
\end{eqnarray}
where $\Gamma(\ )$ is the gamma function, and $E(\ )$ is the complete elliptic integral of the second kind. Using this formula we get $\Omega_0 \approx 2\pi \times 296$ kHz, which also differs from the fitting value by about 100 kHz. We are not clear where these differences come from.

\subsection{Phase-sensitive steady states}

\begin{figure}[h]
\includegraphics[width=10cm]{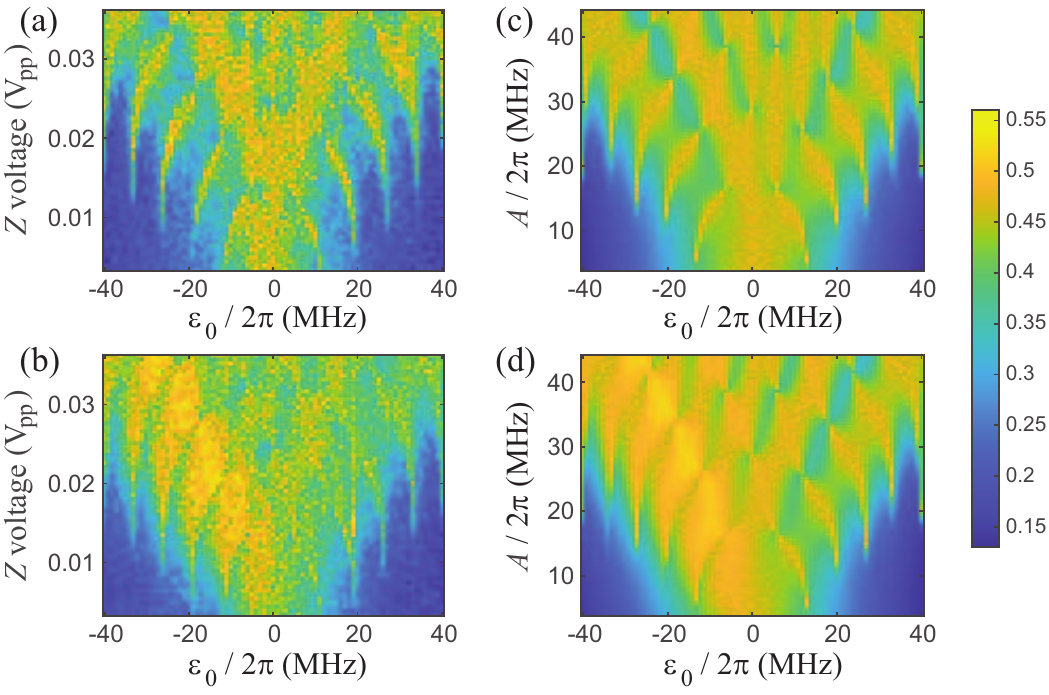}
\caption{
Steady-state $P_{|1\rangle}$ in adiabatic limit. (a) and (b) are experimental data for relative phases $\phi = 44$ degree and $\phi = 134$ degree, respectively. (c) and (d) are numerical results for $\phi = 0.4\pi$ and $\phi = 0.9\pi$, respectively. }
\label{fig:steady}
\end{figure}

To study the phase-dependent steady-state population (LZS spectra), we take $\Omega /2\pi = 12$ MHz, $\omega/2\pi = 6$ MHz, and sweep the $Z$ pulse voltage from 3 mV$_{\rm pp}$ to 36 mV$_{\rm pp}$ which covers the adiabatic limit and the crossover to non-adiabatic limit. 40 $\mu s$ long Rabi pulses are applied to make sure that the qubit is driven to a steady state, and the detuning $\varepsilon_0/2\pi$ is swept from $-40$ MHz to 40 MHz. Figures~\ref{fig:steady}(a) and (c) show the experimental data and the numerical calculation results for $\phi = 0.4\pi$, respectively. A weak dissymmetry can be observed in LZS spectra. By increasing $\phi$ to $0.9\pi$, the dissymmetry becomes more visible, as indicated by Fig.~\ref{fig:steady}(b) and (d).

\section{Conclusion}

We have experimentally and numerically explored the phase-sensitive LZS phenomena using a superconducting Xmon qubit. By choosing such a $dc$ flux bias point, to let the qubit transition frequency near-linearly respond to $ac$ flux modulation, and to let the qubit have a proper dephasing time, we have not only studied time evolution of the qubit population, but also for the first time observed phase dependence in steady-state population. A LZS induced Rabi-like oscillation is observed at the boundary between adiabatic and non-adiabatic limits. Though in this work not much attention has been paid to LZS process in non-adiabatic limit, superconducting Xmon qubit and $IQ$ mixing technique indeed provide a very powerful platform for investigating LZS problems in all parameter regimes.

\appendix

\section{Flux modulation of an Xmon qubit}
For simplicity, we suppose the two Josephson junctions of the Xmon are identical, and limit the flux $-0.5 < \Phi/\Phi_0 < 0.5$. The Hamiltonian of the Xmon reads
\begin{equation}
\hat{H} = -E_J\cos(\pi\Phi/\Phi_0)\cos\hat\gamma + 4E_C \hat{n}^2 ,
\end{equation}
where $\hat\gamma$ and $\hat{n}$ are (dimensionless) phase operator and charge operator, respectively. Defining $\varphi \equiv \pi\Phi/\Phi_0$, this Hamiltonian can be written as
\begin{equation}
\hat{H} = 4E_C\hat{n}^2 + \frac{E_J}{2}\cos(\varphi)\hat\gamma^2 - E_J\cos(\varphi) \left[ \sum_{m=2}^\infty \frac{(-1)^m }{(2m)!} \hat\gamma^{2m} \right] .
\label{eq:H_expand}
\end{equation}
The first two terms on the right-hand-side of Eq.~(\ref{eq:H_expand}) form a simple harmonic oscillator (SHO). Then expressed in terms of SHO creation and annihilation operators, this Hamiltonian is approximately
\begin{equation}
\hat{H}\approx \sqrt{8E_CE_J\cos(\varphi)}\hat{a}^\dag \hat{a} - \frac{E_C}{12}\left( \hat{a}^\dag + \hat{a} \right)^4 .
\label{eq:H_expand_2}
\end{equation}

Now we consider that the flux consists of $dc$ and $ac$ parts, $\varphi = \varphi_{dc} + \varphi_{ac}(t)$, and expand $\cos(\varphi)$ to the leading order of $\varphi_{ac}(t)$,
\begin{equation}
\cos(\varphi) \approx \cos(\varphi_{dc}) - \varphi_{ac}(t)\sin(\varphi_{dc}) .
\end{equation}
Therefore the Hamiltonian Eq.~(\ref{eq:H_expand_2}) can be written as
\begin{equation}
\hat{H}\approx \hbar\omega_p\sqrt{1-\varphi_{ac}(t)\tan(\varphi_{dc})} \hat{a}^\dag \hat{a} - \frac{E_C}{12}\left( \hat{a}^\dag + \hat{a} \right)^4 ,
\end{equation}
where $\omega_p = \sqrt{8E_CE_J\cos(\varphi_{dc})}/\hbar$. By making the approximation
\begin{equation}
\sqrt{1-\varphi_{ac}(t)\tan(\varphi_{dc})} \approx 1-\frac{\tan(\varphi_{dc})}{2}\varphi_{ac}(t) ,
\end{equation}
we can rewrite the Hamiltonian as
\begin{equation}
\hat{H}\approx \hbar\omega_p \hat{a}^\dag \hat{a} - \frac{E_C}{12}\left( \hat{a}^\dag + \hat{a} \right)^4 - \frac{\hbar\omega_p}{2}\tan(\varphi_{dc})\varphi_{ac}(t) \hat{a}^\dag \hat{a} .
\label{eq:H_expand_3}
\end{equation}

Assuming the $ac$ flux has the form $\varphi_{ac}(t) = \varphi_{ac}\cos(\omega t + \phi)$, and truncating the Hamiltonian Eq.~(\ref{eq:H_expand_3}) to its lowest two energy levels, we get
\begin{equation}
\hat{H} = -\frac{\hbar}{2}\left[ \omega_0 - A \cos(\omega t + \phi) \right]\hat\sigma_z ,
\end{equation}
with $\omega_0 = \omega_p - E_C/\hbar$, and $A = \omega_p\varphi_{ac}\tan(\varphi_{dc})/2$.

\section{Calibrations of parameters}

\begin{figure}
\includegraphics[width=7.5cm]{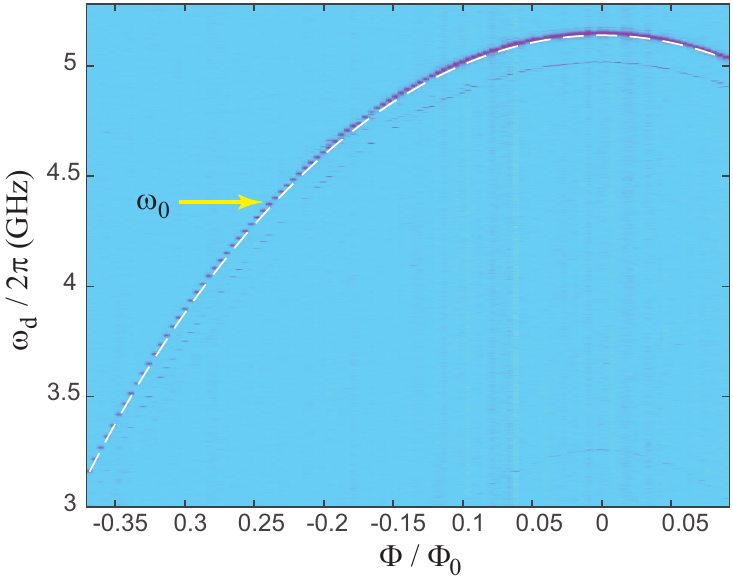}
\caption{
Qubit transition frequency versus flux. The $|0\rangle \rightarrow |1\rangle$ transition frequencies are fitted (the white dashed curve) by $E_C/h = 0.264$ GHz and $E_J/h = 13.822$ GHz. }
\label{fig:qubit_spectro}
\end{figure}

Figure~\ref{fig:qubit_spectro} shows the Xmon spectra with only $dc$ flux bias. Due to the relatively strong $XY$ drive amplitude used in spectroscopy, the two-photon transition process coupling Xmon's ground state $|0\rangle$ and second excited state $|2\rangle$ is observed (the less visible parallel curve under the main spectra curve). The frequency difference (132 MHz for this sample) between this two-photon transition and the single-photon $|0\rangle$ to $|1\rangle$ (Xmon's first excited state) transition is approximately $E_C/(2h)$ \cite{koch07}. By fitting the $|0\rangle$ to $|1\rangle$ transition (qubit) frequencies with Eq.~(\ref{eq:omega_10}), as shown by the white dashed curve in Fig.~\ref{fig:qubit_spectro}, we obtain the Josephson energy of this Xmon $E_J = h \times 13.822$ GHz.

\begin{figure}
\includegraphics[width=7.5cm]{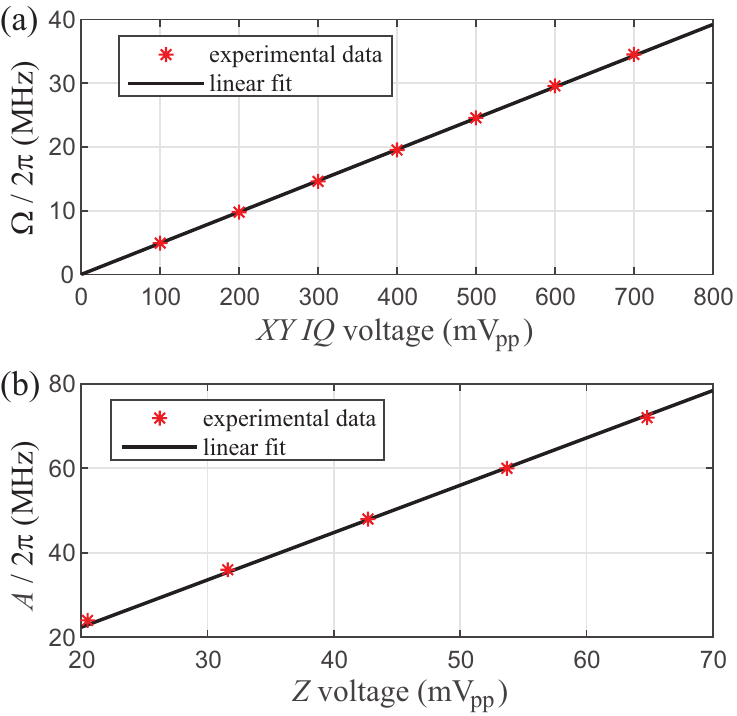}
\caption{
Calibrations of (a) Rabi frequency $\Omega$ versus $XY\ IQ$ pulses peak-to-peak voltage, and (b) $Z$ modulation amplitude $A$ versus $Z$ pulse peak-to-peak voltage. }
\label{fig:calib}
\end{figure}

Figure~\ref{fig:calib}(a) shows a calibration of Rabi frequency $\Omega$ for different $XY\ IQ$ pulses peak-to-peak voltages at qubit frequency 4.365 GHz. The $Z$ modulation frequency $\omega$ is precisely known in experiments, but the modulation amplitude $A$ of qubit transition frequency can not be directly measured. In small coupling regime $\Omega\ll \omega$, the qubit $|1\rangle$ state population is approximately zero if the $n$th Bessel function of the first kind $J_n(A/\omega)$ equals zero. Therefore the modulation amplitude $A$ for different $Z$ pulse peak-to-peak voltage can be calibrated using, e.g. the first zero of $J_0(A/\omega)$, at which $A\approx 2.4\omega$. To do this calibration, we set detuning $\varepsilon_0 = 0$ and Rabi frequency $\Omega/2\pi = 2.5$ MHz, then find $Z$ pulse peak-to-peak voltages corresponding to zero population for various modulation frequencies $\omega/2\pi = 10, 15, 20, 25, 30$ MHz. The calibration result is shown in Fig.~\ref{fig:calib}(b).

\acknowledgements
Enlightening and fruitful discussions with Professor Barry Garraway at University of Sussex is gratefully acknowledged.

Project supported by the Key-Area Research and Development Program of Guangdong Province (Grant No. 2018B030326001), the National Natural Science Foundation of China (Grant No. U1801661, No. 11874065, and Youth Project No. 11904158), the Guangdong Provincial Key Laboratory (Grant No. 2019B121203002), the Natural Science Foundation of Hunan Province (Grant No. 2018JJ1031), and the Science, Technology and Innovation Commission of Shenzhen Municipality (Grants No. JCYJ20170412152620376, No. YTDPT20181011104202253).

\end{document}